# Fast library-driven approach for implementation of the voxel spread function technique for correcting magnetic field inhomogeneity artifacts

Jie Wen[1], Feiyan Zeng[1], Dmitriy Yablonskiy[2], Alexander Sukstansky[2], Ying Liu[1], Bin Cai[3], Yong Zhang[4], Weifu Lv[1]

1. Radiology, The First Affiliated Hospital of USTC, Division of Life Sciences and Medicine, University of Science and Technology of China, Hefei, Anhui, China, 230001
2. Radiology, Washington University in St. Louis, St. Louis, MO, USA, 63110
3. Radiation Oncology, Washington University in St. Louis, St. Louis, MO, USA, 63110
4. GE Healthcare, Shanghai, China, 201203

Corresponding author:

Jie Wen, PhD

Email: jiewen@ustc.edu.cn

Tel: +86-0551-62283486

Radiology, The First Affiliated Hospital of USTC, Division of Life Sciences and Medicine, University of Science and Technology of China, Hefei, Anhui, China, 230001

Key words: voxel spread function, library, fast calculation, R2*, gradient echo

# Abstract

*Purpose*: Previously-developed Voxel Spread Function (VSF) method (Yablonskiy, et al, MRM, 2013;70:1283) provides means to correct artifacts induced by macroscopic magnetic field inhomogeneities in the images obtained by multi-Gradient-Recalled-Echo (mGRE) techniques. The goal of this study is to develop a library-driven approach for fast VSF implementation.

*Methods*: The VSF approach describes the contribution of the magnetic field inhomogeneity effects on the mGRE signal decay in terms of the F-function calculated from mGRE phase and magnitude images. A pre-calculated library accounting for a variety of background field gradients caused by magnetic field inhomogeneities was used herein to speed up calculation of the F-function and to generate quantitative R2* maps from the mGRE data collected from two healthy volunteers.

*Results*: As compared with direct calculation of the F-function based on a voxel-wise approach, the new library-driven method substantially reduces computational time from several hours to few minutes, while, at the same time, providing similar accuracy of R2* mapping.

*Conclusion*: The new procedure proposed in this study provides a fast post-processing algorithm that can be incorporated in the quantitative analysis of mGRE data to account for background field inhomogeneity artifacts, thus can facilitate the applications of mGRE-based quantitative techniques in clinical practices.

# 1. Introduction

It is well-known that macroscopic $B0$ magnetic field inhomogeneities adversely affect quantitative measurements of relaxation parameters obtained with multi-Gradient-Recalled-Echo (mGRE) MRI (1) if not accounted for.

It was demonstrated recently that a voxel spread function (VSF) method of correcting field inhomogeneities (2) incorporated in a quantitative mGRE-based approach (3) provides quantitative information on brain tissue cellular structure (4) as well as its alterations under healthy aging (4) and different neurological diseases (5-8). The VSF method takes advantage of utilizing both magnitude and phase of mGRE data from a single scan, hence it saves acquisition time and avoids misregistration errors. However, a direct implementation of the VSF method requires a large number of voxel-dependent convolutions resulting in long computational time.

In this study, we propose a library-driven approach for fast implementation of the VSF method. Instead of the direct voxel-wise calculation, a pre-calculated library accounting for a variety of background field gradients caused by magnetic field inhomogeneities is used. The library-driven approach substantially reduces the computational time of VSF from hours to a few minutes, and provides similar accuracy compared with the direct calculation. This can facilitate the applications of mGRE-based quantitative techniques in clinical practices.

# 2. Methods

## 2.1 Brief Overview of the VSF method

As described in Ref. (8), the image of mGRE-MRI can be expressed as:

$$S_n(TE) = \sum_m \Psi_{nm}(TE) \cdot \sigma_m(TE) \qquad [1]$$

where $S_n(TE)$ is the complex images, $\sigma_m(TE)$ is the "ideal" signal from the $m^{th}$ voxel that is free from $B0$-field inhomogeneity artifacts. $\Psi_{nm}(TE)$ is a voxel-dependent convolution kernel, which in the 3D case can be presented as:

$$\Psi_{nm}(TE) = \exp(i\omega_m \cdot TE + i\varphi_0) \cdot \eta_x \cdot \eta_y \cdot \eta_z \qquad [2]$$

where the voxel indices $n$ and $m$ are considered as 3D vectors: $(n_x, n_y, n_z)$ and $(m_x, m_y, m_z)$, $\omega_m$ is the frequency shifts. $\varphi_0$ is the signal phase shift at $TE=0$, mainly resulted from RF field inhomogeneities. The factors $\eta_j$ ($j = x, y, z$) have the form:

$$\eta_j = \sum_{q=1}^{N_q} \text{sinc}(q - q_{mj}(TE)) \cdot \exp(2\pi i q(n_j - m_j)) \qquad [3]$$

where $q = \{-1/2, -1/2 + 1/N_q, ..., 1/2 - 1/N_q\}$ is the 1D normalized $k$-space. For images obtained with the Hanning filter, Eq. [3] should be modified by adding an additional factor $\cos^2(\pi q)$.

The parameters $q_{mj}$ describe phase dispersion across the $m^{th}$ voxel:

$$q_{mj}(TE) = \frac{1}{2\pi}(g_{mj} \cdot TE + G_{mj}) \qquad [4]$$

where $\mathbf{g}_m$ and $\mathbf{G}_m$ are the gradients for the $m^{th}$ voxel calculated from the spatially unwrapped frequency $\omega$ and phase $\varphi_0$, respectively. In the 2D case, the factor $\eta_z = \text{sinc}(q_{mz}(TE))$.

Using the similarity approximation,

$$\sigma_m(TE) \Rightarrow \sigma_n(TE) \frac{|S_m(0)|}{|S_n(0)|} \qquad [5]$$

Eq. [1] reduces to

$$S_n(TE) = \sigma_n(TE) \cdot F_n(TE) \qquad [6]$$

where the *F*-function, describing the influence of *B*0 field inhomogeneities on images, is given by

$$F_n(TE) = \frac{1}{|S_n(0)|} \sum_m \Psi_{nm}(TE) \cdot |S_m(0)| \qquad [7]$$

In practice, the sum in Eq. [7] is taken over several voxels adjacent to a given voxel *n*. Even for few neighbors included in this sum (see (10)), a large number of voxel-dependent convolutions leads to long computational time. For a typical head scan using mGRE sequence with a matrix size of 256×256×72×10, the calculation time of the *F*-function in Matlab using 4 neighbors (in each direction) could take several hours on a computer with typical configurations.

## 2.2 Library-driven approach

In the voxel-wise direct calculation, the factors $\eta$ are calculated independently for each voxel. As mentioned above, this procedure usually costs long computational time. In this study, we propose a new way to substantially accelerate the calculation by using a pre-calculated library for $\eta$. As shown in Eq. [3], the factors $\eta$ depend on three parameters: the phase dispersion $q_m$, neighbor index $(n-m)$, and $N_q$. When phased-array coils and channel combination methods (4) are used, the parameter $q_m$ depends only on the frequency gradient $g_m$. In the human brain, the maximum $g_m$ is well below 1500 rads/voxel. When using an accuracy of 0.1 rads/voxel, a total number of 30,000 points is sufficient to build the library for $\eta$. Although $\eta$ also depends on the number of $q$-points in the sum in Eq. [3], our simulation shows that as long as using a sufficient large $N_q \geq 256$, the factor $\eta$ is fairly stable (see **Figure 1**).

Since changing the sign of either $q_m$ or neighbor index $(n-m)$ will result in a complex conjugate of $\eta$, one can only use non-negative values of $q$ and non-negative neighbor index $(n-m)$ for calculations. This can further reduce the computational time for library building. This library of $\eta$ is then used for *F*-function calculation: for each voxel and its neighbors with particular $g_m$ values, corresponding values of $\eta$ can be readily found from the library. Importantly, replacing the direct computation by this new library-driven approach substantially reduces the computing time from hours to a few minutes. As long as including sufficiently large $N_q$ and number of neighboring voxels, the library only needs to be built once and can be used for future calculations.

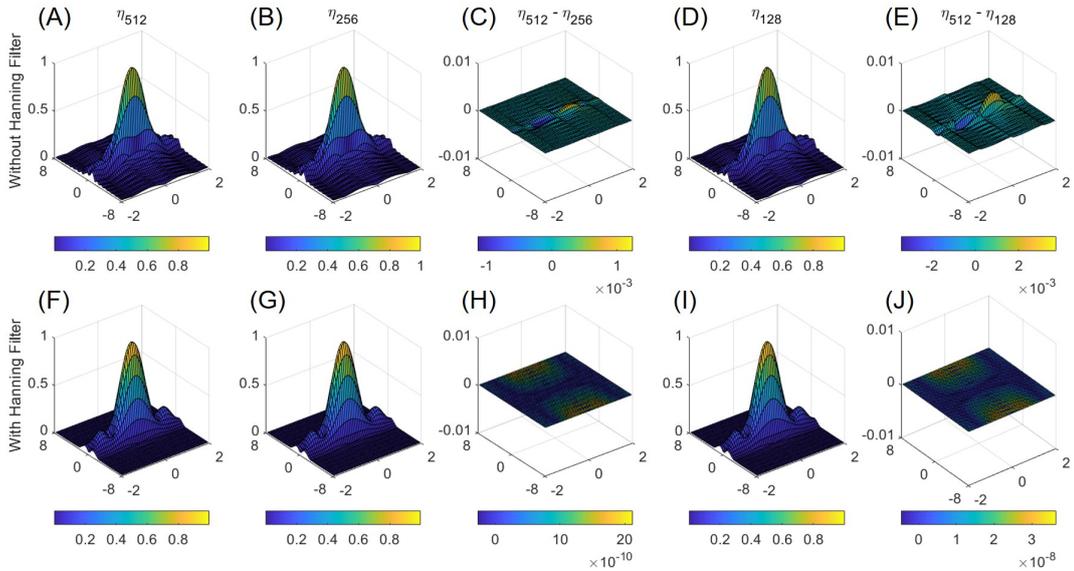

**Figure 1** Simulation of $\eta$ as a function of the distance from a given voxel to the neighboring voxel (-8 to 8) and the parameter $q_m \in (-2, 2)$. A-E: $\eta$ for the non-filtered

data; F-J: $\eta$ for the Hanning filtered data. The factor $\eta$ calculated with $N_q = 512$ (A and F), $N_q = 256$ (B and G) and $N_q = 128$ (D and I) points are shown. The corresponding differences are shown in C, E, H and J. It turns out that when using a sufficient large number of $q$-points in the sum in Eq. [3] ($N_q \geq 256$), the changes of the values of $\eta$ are small ($< 1.5 \times 10^{-3}$ for the non-filtered data and $< 2.5 \times 10^{-10}$ for the Hanning filtered data).

## 2.3 Image acquisition and processing

Images were collected from two healthy volunteers (both are 26 years old female) using a 3T MRI scanner (GE Discovery 750w) equipped with a 12-channel phased-array head coil. All studies were approved by the IRB of The First Affiliated Hospital of USTC. All the participated healthy volunteers have provided written consents.

High resolution datasets with a voxel size of 1×1×2 mm³ were acquired using a three dimensional (3D) mGRE sequence (ESWAN on GE scanners) with a flip angle of 20° and TR=50.4 ms. For each acquisition, 10 echoes were collected with first echo time TE1 = 3.4 ms and echo spacing ΔTE = 3.8 ms. The total acquisition time was around 11 min. The complex images from different channels were combined using a previously introduced method (4,11). The combined mGRE datasets were then used for calculation of the $F$-function and transverse relaxation rate constant (R2*). The libraries were built for 3 different values of the parameter $N_q$ (512, 256, 128). The $F$-function was calculated with 4 neighbors for Hanning-filtered data and 16 neighbors for non-filtered data. A maximum value of 1500 rads/voxel and a step of 0.1 rads/voxel were used for the gradient $g_m$. All computations were done in Matlab2019a (The MathWorks, Inc., Natick MA, USA) on a computer with standard configurations: 3.2 GHz CPU and 16 GB RAM.

## 3. Results

**Figure** 2 shows the comparisons of fitted R2* maps without and with the *F*-function. Similarly to (12), the results show that the *F*-function significantly reduces magnetic field inhomogeneity artifacts on R2* maps, as indicated by red arrows in **Figure** 2.

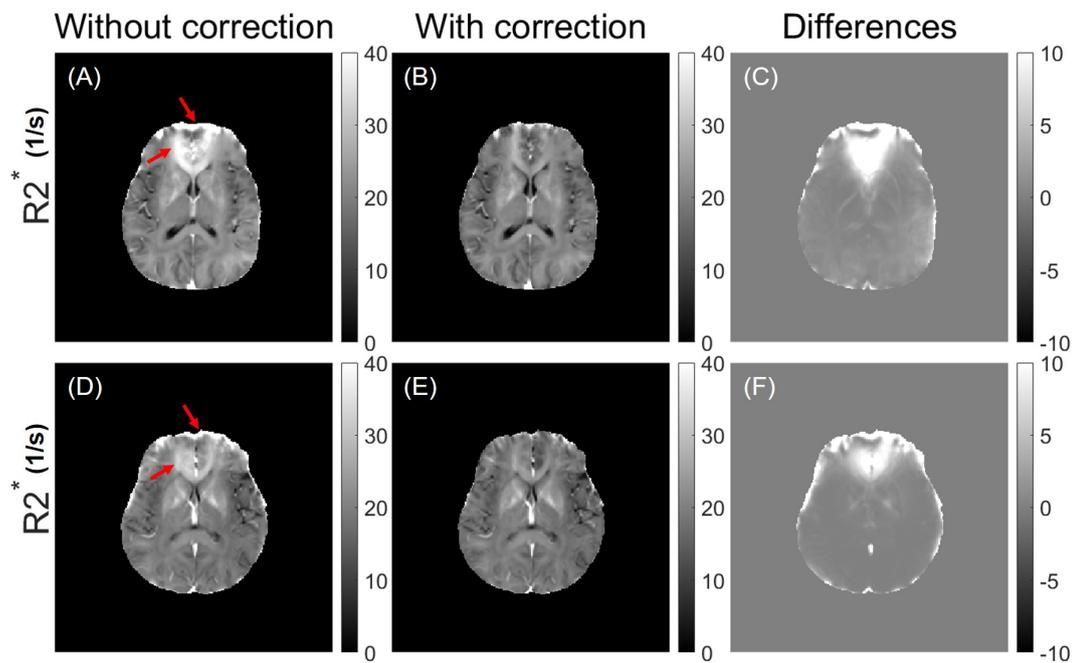

**Figure** 2 Examples of R2* maps calculated without (A and D) and with (B and E) F-function and their differences (C and F) for two healthy volunteers (both are 26 years old female). Regions substantially affected by the *B*0 field inhomogeneity are indicated by red arrows. The Hanning-filtered data were used for R2* calculation.

The comparison of the *F*-functions calculated using direct voxel-wise calculation and library-driven approach are shown in **Figure** 3. The direct calculation of the *F*-function required about 5.5 hours on a computer with a 3.2 GHz CPU and 16 GB RAM, while the new method requires only 3 minutes on the same computer. The differences are well below $10^{-4}$ for the *F*-function and $10^{-3}$ for R2*, respectively (panels C, F).

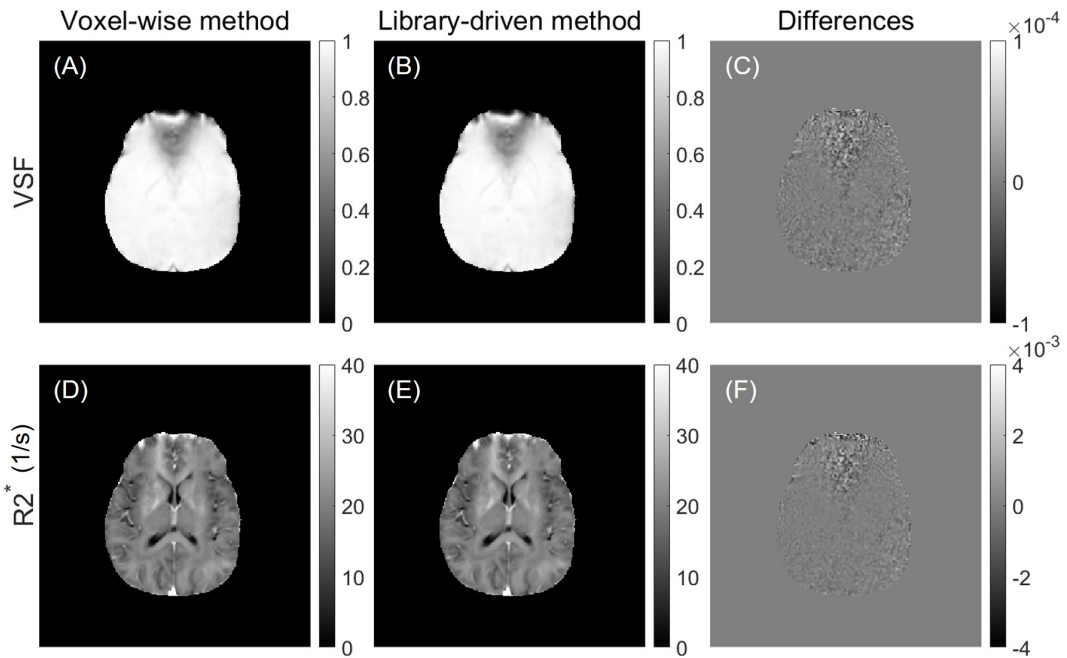

**Figure 3** Comparison of the *F*-function at TE = 37.5 ms (A-C) and R2* maps (D-F) using voxel-wise method (A and D) and new library-driven method (B and E). The differences are shown in C and F. The Hanning-filtered images were used to calculate the F-function and R2* maps.

**Figure 4** compares R2* maps calculated using the *F*-function with different numbers of neighbors included in the sum in Eq. [7]. As demonstrated previously (4), using 4 neighbors (in each direction) is sufficient to get accurate R2* values from Hanning-filtered data, while for non-filtered data, more neighbors (16 at least) are required.

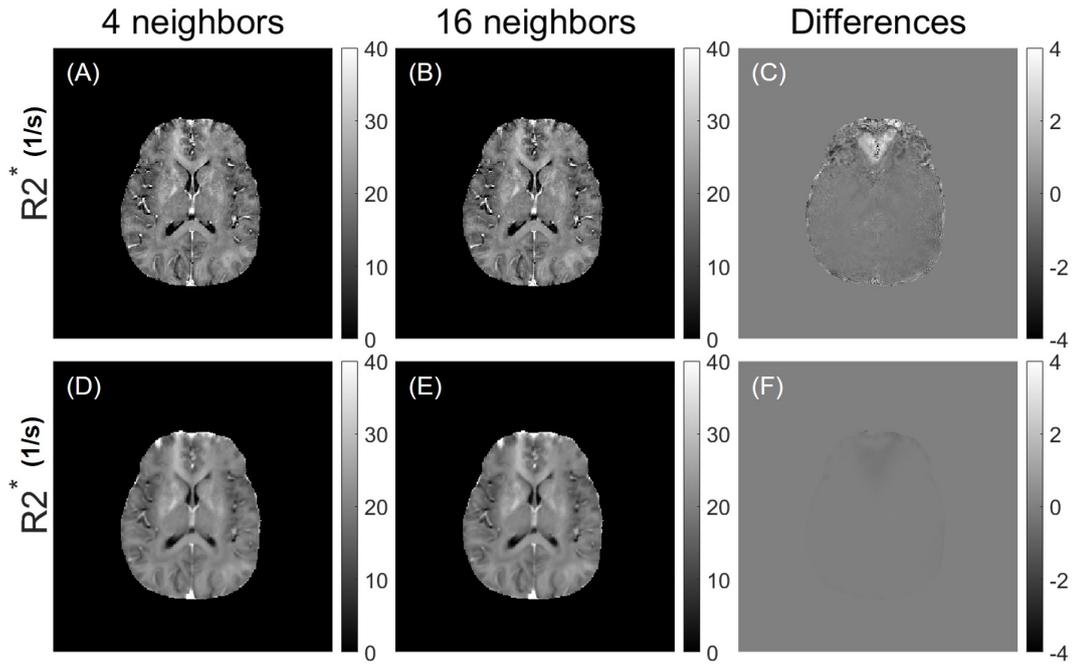

**Figure 4** Comparison of R2* maps calculated using the *F*-function with different number of neighboring voxels included in the sum in Eq. [7]. Results for non-filtered (A-C) and Hanning-filtered (D-F) data are shown.

R2* maps calculated using the *F*-function with different number of *q*-points $N_q$ are shown in **Figure 5**. Accurate R2* calculations are guaranteed when sufficient large number of *q*-points in Eq. [3] ($N_q \geqslant 256$) are employed, which coincides with the simulation results.

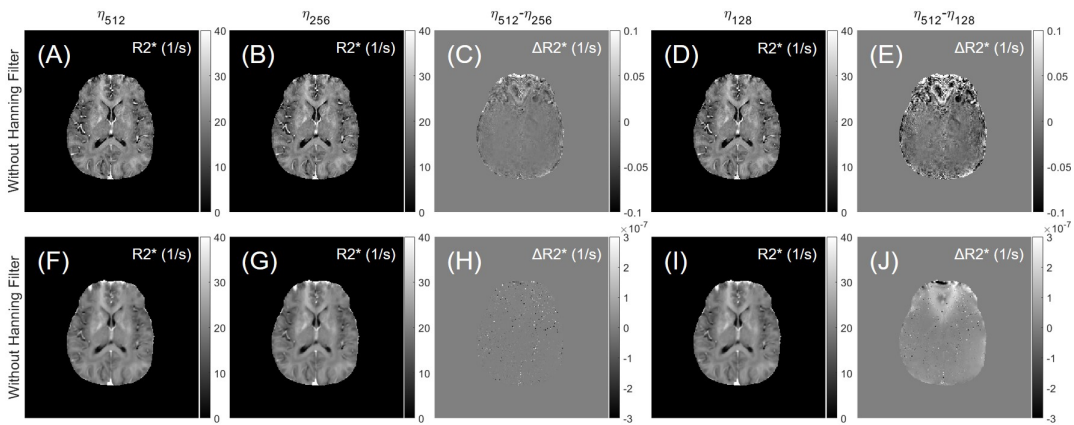

**Figure 5** Comparison of R2* maps calculated using F-function with different number of

$q$- points $N_q$ (A and F: 512 points; B and G: 256 points; D and I: 128 points). R2* and differences ($\Delta$R2*) maps were calculated from non-filtered (A-E) and Hanning-filtered (F-J) data.

## 4. Discussion and Conclusions

Macroscopic $B0$ magnetic field inhomogeneities adversely affect MRI images, especially the ones based on a gradient echo acquisition. They not only lead to image distortions (12), but also degrades quantitative measurements of relaxation parameters (4,13-16) if not accounted for. Numerous methods have been proposed to correct for these adverse effects, including correcting methods during data collection (16-23) and post processing approaches (1,3,24-26). In this paper we focus on the VSF method (27) that allows accounting for the effects of magnetic field inhomogeneities in data obtained with mGRE sequences.

Quantitative parameters that are evaluated using mGRE-based techniques, such as R2* mapping, are usually suffered from magnetic field inhomogeneity artifacts. These artifacts are extremely severe in those regions that are close to the tissue-air interfaces in sinuses, as shown in **Figure** 2. These artifacts often corrupt the quantitative measurements, thus hinder the clinical application of these quantitative techniques. The VSF method was proposed previously to correct for this effect, however, direct voxel-by-voxel calculations of the $F$-function requires long computational time. In this study, we presented a library-driven approach for fast VSF implementation. By using a pre-calculated library, the calculation time of the $F$-function has been reduced from hours to a few minutes.

We showed through simulations (**Figure 1**) and experiments (**Figure 3**) that by selecting a sufficient number of $q$-points ($N_q$) in the sum in Eq. [3] and the number of neighboring voxels included in the sum in Eq. [7] to build the library, the new library-driven method substantially reduces computational time while, at the same time, provides similar accuracy compared with the direct voxel-wise calculation. The reduction of computational time makes it feasible to implement the VSF method with more neighbors included the sum in Eq. [7], which is necessary for non- filtered data.

In conclusion, we presented in this Note a library-driven approach for fast VSF method implementation. The new procedure provides a fast way to correct mGRE-based images for background field inhomogeneity artifacts, thus can facilitate the applications of mGRE-based quantitative techniques in clinical practices.

## Acknowledgments

This work was supported by "the Fundamental Research Funds for the Central Universities" WK9110000119.